\documentclass[12pt]{iopart}
\usepackage{graphicx}

\begin{document}

\title{Double core-hole spectroscopy of transient plasmas produced in the interaction
of ultraintense x-ray pulses with neon}


\author{Cheng Gao$^{1}$, Jiaolong Zeng$^{1}$, and Jianmin Yuan$^{1,2}$}


\address{$^1$Department of Physics, College of Science,
National University of Defense Technology, Changsha Hunan 410073, P. R. China}
\address{$^2$IFSA Collaborative Innovation Center, Shanghai Jiao Tong University,
Shanghai 200240, P. R. China}

\ead{jlzeng@nudt.edu.cn, jmyuan@nudt.edu.cn}


\begin{abstract}

Double core-hole (DCH) spectroscopy is investigated systematically for neon
atomic system in the interaction with ultraintense x-ray pulses with photon energy
from 937 eV to 2000 eV. A time-dependent rate equation, implemented in the detailed
level accounting approximation, is utilized to study the dynamical evolution of the
level population and emission properties of the highly transient plasmas. For x-ray pulses
with photon energy in the range of 937-1030 eV, where $1s\rightarrow 2p$ resonance
absorption from single core-hole (SCH) states of neon charge states exist,
inner-shell resonant absorption (IRA) effects play important roles
in the time evolution of population and DCH spectroscopy. Such IRA physical
effects are illustrated in detail by investigating the interaction of x-ray pulses
at a photon energy of 944 eV, which corresponds to the $1s\rightarrow 2p$ resonant
absorption from the SCH states ($1s2s^22p^4$, $1s2s2p^5$ and
$1s2p^6$) of Ne$^{3+}$. After averaging over the space and time distribution of the
x-ray pulses, DCH spectroscopy at photon energies of 937, 944, 955, 968, 980,
and 990 eV (corresponding to the $1s\rightarrow 2p$ resonance energies of
Ne$^{2+}$-Ne$^{7+}$, respectively) are studied in detail. The complex mechanisms
of producing the DCH states are discussed, taking the DCH spectroscopy at 937 and
980 eV as examples. For photon energy larger than 1362 eV, there are no resonance
absorption in the interaction, resulting in a similar physical picture of DCH
spectroscopy. The dominant DCH states are due to higher ionization stages of
Ne$^{7+}$-Ne$^{9+}$, while the DCH spectroscopy from lower charge states Ne$^{2+}$-Ne$^{6+}$
are much weaker. With the increase of x-ray photon energy, the contributions
from the lower charge states become larger.

\end{abstract}

\maketitle


\section{Introduction}

Production of double core-hole (DCH) states represents a quite recent experimental
achievement by using synchrotron radiation and x-ray free-electron lasers
(XFELs) \cite{Piancastelli}. Two main experimental methods have been used to induce
such highly transient quantum states with a lifetime of $\sim$1 femtosecond (fs).
One is by using synchrotron radiation and then single photon inner-shell
photoionization occurs followed by the simultaneous ejection of two core electrons.
The other is by using XFELs and multiple-photon absorption, resulting in the
production of DCH states via sequential absorption of two x-ray photons on a time
scale of a few fs. The probability of the former method is in general much smaller
than the latter by using ultraintense x-ray pulses. As is well known, the intensity
of x-ray radiation produced by synchrotron radiation is smaller than that of XFELs
by more than six orders of magnitudes and hence creation of DCH states by sequential
photoionization was not
possible. However, such a case is changed by the ultraintense XFEL pulses
such as the Linac Coherent Light Source (LCLS) \cite{Emma} and the Spring-8
Angstrom Compact free electron LAser (SACLA) \cite{Ishikawa}.
With the new generation light source, hollow atoms and molecules are readily
produced \cite{Bostedt,Young,Berrah,Fang,Cryan,Salen,Tamasaku,Frasinski}.

DCH spectroscopy carries on rich information on the transient plasmas created
in the interaction of x-ray pulses with matters (atoms, molecules, clusters, and
solid-state matter) and therefore it has practical
applications in many aspects. Compared with conventional inner-shell photoelectron
spectroscopy, it has unique features in the investigations of such highly transient
plasmas \cite{Santra}. Auger electrons emitted from DCH states are
a distinctive signature of hollow atom formation and hence electron spectra
can act as the evidence of hollow atoms.
Recently, investigations carried out by Berrah {\it et al.} \cite{Berrah},
Salen {\it et al.} \cite{Salen} and Takahashi and Ueda \cite{Takahashi} experimentally
verified that DCH spectroscopy provides a powerful means for chemical analysis.
It has shown that it has greater potential to probe the local chemical environment
than by detecting single core-hole (SCH) spectroscopy.
Secondly, combining with accurate theoretical simulations \cite{Ueda}, DCH
spectroscopy provides a useful tool for estimating or even diagnosing the intensity
and duration of x-ray pulses \cite{Larsson}.
In dense plasmas, it reflects the screening effects on atoms embedded in the plasmas
produced by x-ray laser pulses \cite{Vinko,Ciricosta,Rackstraw,Preston}
and thus it can be utilized to deduce the environmental effects.

Neon is a prototype object to investigate the interaction with high-intensity and
short duration x-ray pulses \cite{Young,Doumy,Rohringer,Kanter,Ciricosta2,Nikolopoulos}.
However, detailed and systematic investigations of the DCH spectroscopy on this object
are lacking very much.
Young {\it et al.} \cite{Young} experimentally observed the existence of DCH
states of neon at a photon energy of 1050 eV by measuring the electron spectra
at the direction of incident x-ray polarization. To the best of our knowledge,
no measurements are available for the emission spectra produced by the DCH
states of neon. Our understanding on
the DCH states and spectroscopy for neon interacting with XFELs is rather limited
and incomplete. It is worthwhile to systematically study it over a wide range
of photon energy.

In this work, we systematically investigate the emission properties of DCH states
of transient neon plasmas created in the interaction of XFELs with photon energy
ranging from 937 eV to 2000 eV.
A time-dependent rate equation (TDRE) approach \cite{Ciricosta2,Rohringer2,Abdallah}
based on the collisional-radiative model is employed to study the evolution
dynamics of charge state distribution (CSD) and emission properties.
A detailed level accounting (DLA) model is utilized to describe the
quantum states of the neon system. The DLA model has been proved to be one of the
most accurate method in the computation of the radiative opacity of hot dense
plasmas in local thermodynamic equilibrium \cite{Zeng,Zeng2,Zeng3,Gao}.
It is also one of the most accurate method to describe
the XFELs interaction with atoms. Based on the TDRE and DLA approaches,
we systematically investigated the DCH spectroscopy for the XFELs interaction
with neon at different photon energy over a wide range. For the x-ray pulses,
we assume Gaussian profiles for
the spatial, time and frequency distribution.

\section{Theoretical methods}

In the treatment of the interaction of XFELs with atoms, we utilized a DLA formalism,
where the quantum states of our concerned physical system are accurate to
fine-structure level. The population $n_i$
of the fine-structure level $i$ is obtained by solving the rate equation
\begin{equation}
\frac{dn_i}{dt}=\sum\limits_{j\neq
i}^{N_L}n_jR_{ji}-n_i\sum\limits_{j \neq i}^{N_L}R_{ij},
\end{equation}
where $R_{ij}$ and $R_{ji}$ represent the rates which depopulate and populate between
the levels $i$ and $j$ and $N_L$ is the total number of levels
included in the calculation. The levels of all ionization stages from the
neutral to fully stripped neon ions are included up to principal quantum number
$n$=4 with SCH and DCH states being taken into account. The rates connecting different
levels include all main
microscopic atomic processes due to photons and electrons, namely
photo-excitation, photoionization, electron impact excitation,
electron impact ionization, Auger decay and their inverse processes \cite{Gao1,Gao2}.

The rate due to photons (radiation field) is determined by the cross sections
of corresponding microscopic process \cite{Xiang,Rose}
\begin{equation}
R_{ij}(r,t)=\int\frac{I(r,t,h\nu)}{h\nu}\sigma_{ij}(h\nu)d(h\nu),
\end{equation}
where $I(r,t,h\nu)$ is the XFEL intensity at space position $r$ from the center
of the laser spot, time $t$ and the photon frequency $\nu$, and
$\sigma_{ij}(h\nu)$ is the photoexcitation or photoionization cross
section at photon energy $h\nu$.
The intensity of XFELs is assumed to have a Gaussian profile
on the space, time and photon frequency
\begin{equation}
I(r,t,h\nu)=I_0e^{-\ln2(\frac{r}{\Delta})^2}e^{-\ln2(\frac{t}{\tau})^2}
\sqrt{\frac{\ln2}{\pi\Gamma^2}}e^{-\ln2(\frac{h\nu-h\nu_0}{\Gamma})^2},
\end{equation}
where $I_0$ is the peak intensity and
$h\nu_0$ is the photon energy of the x-ray pulse.
$\Delta$, $\tau$ and $\Gamma$ are the half width at half maximum (HWHM)
of Gaussian profile of the x-ray pulse on the distribution of space, time
and photon energy, respectively. Note that the dependence of intensity
on photon energy (frequency) has been multiplied by a constant quantity
to satisfy the normalization (to 1). The energy of x-ray pulses
can be obtained by integrating the intensity over the space, time and photon energy
over all possible variable regions. Such a quantitative connection of
pulse energy and intensity can help us to better understand the physics
of the interaction.
The initial density of neon is taken to be 1.0$\times$10$^{17}$ cm$^{-3}$
in this work and thus the rates due to electrons are small compared with those of
photons and Auger decay rates \cite{Ciricosta2}.
The details on the atomic model and atomic data can be found elsewhere \cite{Gao1}.
Briefly, we include the quantum states of single and double excitation states
including core-hole ($1s$) states up to
principal quantum number $n$=4 for all ionization stages from the neutral
atom to bare ion in the rate equation. The required atomic data including the
spontaneous radiative decay rate, photoionization and photo-excitation cross
section, and Auger decay rate connecting all these levels are obtained by solving
a relativistic Dirac equation \cite{Gu}.
The direct double Auger decay rates are obtained by using simplified formulas
according to the knock-out (KO) and shake-off (SO) mechanisms
as we demonstrated \cite{Zeng4,Zeng5,lpf1,Gao3}.

After solving the rate equation, we obtain the level populations which can be
used to calculate the emission properties of plasmas.
The emissivity $j(h\nu)$ at photon energy $h\nu$ due to DCH states reads as
\begin{equation}
j(h\nu)=\sum\limits_{j>i}n_jh\nu A_{ji}S(h\nu),
\end{equation}
where $n_j$ is the population of the upper level of the transition
$j\rightarrow i$, $A_{ji}$ and $S(h\nu)$ are the radiative transition probability
and line profile of the transition $j\rightarrow i$, respectively. $S(h\nu)$ is
taken to be a Lorentian profile to account for both the natural lifetime
broadening and autoionization width. If we only consider the DCH spectroscopy,
we limit the summation on the upper levels which are DCH states.

\section{Results and discussions}

To produce DCH states, the lowest photon energy should be high enough to ionize
or excite two $1s$ electrons of the neutral neon. As we know, the ionization
potential (IP) will increase for the $1s$ electron from the SCH state compared to
IP from levels of ground configuration, hence the way with lowest energy should
be photoionizing the first $1s$ electron and then photo-exciting another $1s$ electron
to $2p$. To have a quantitative understanding,
we give the single and double ionization potential (SIP and DIP) of $1s$ electrons
from the atomic to hydrogen-like neon in table 1. Note that hydrogen-like neon
has only one electron and thus there is no DIP for it. The experimental values
are given wherever available for atomic Ne \cite{Thomas2}, Ne$^{4+}$-Ne$^{6+}$ \cite{Bruch}
and Ne$^{8+}$ \cite{NIST}. For those ions where no experimental results
are available, our theoretical results are given.
From the inspection of table 1, we see that the $1s$ SIP of atomic neon
is 870.3 eV \cite{Thomas2} and the lowest DIP is 986.0 eV for Ne$^{2+}$
(production of $1s^02s^22p^6$).
It seems that the lowest photon energy which can produce DCH states is 986.0 eV.
This conclusion is correct if one only considers the direct photoionization channels.
However, if we include the resonance absorption from the SCH states,
the above physical picture will be modified. To have an overall view of the
resonance absorption of $1s\rightarrow2p$, we show the absorption cross sections
from all possible energy levels belonging to the configurations of $1s2s^m2p^n$
($m$+$n$=8-0, corresponds to Ne$^{1+}$-Ne$^{9+}$, respectively) in Fig. 1.
The result is obtained by summing the photoexcitation cross section of
resonance absorption $1s\rightarrow2p$ from all levels of $1s2s^m2p^n$
of all possible charge states
\begin{equation}
\sigma(h\nu)=\sum\limits_{j>i}\frac{\pi he^2}{m_ec}f_{ij}S(h\nu),
\end{equation}
where $h$ is the Planck constant, $m_e$ is the electron mass, $S(h\nu)$ is
the line profile. A Lorentian profile is assumed with natural lifetime
broadening and autoionization width being taken into account.

From the inspection of Fig. 1, we can determine that the lowest photon energy
which can produce DCH states is $\sim$937.0 eV, which is the resonance energy
of $1s2s^22p^5\rightarrow1s^02s^22p^6$ for Ne$^{2+}$. Photons with this energy can first
effectively ionize atomic neon and Ne$^{1+}$ and then produce DCH state
$1s^02s^22p^6$ (Ne$^{2+}$) by resonance photo-excitation process. Actually, there are
many resonance absorptions from different ionization stages in the photon energy
range 935.0 eV-1030.0 eV. First, we point out that the resonance absorption
for x-ray photons with an energy in this range plays an important role
in the evolution dynamics of population and DCH
spectroscopy just as we demonstrated previously \cite{Xiang}.
We take the x-ray pulse with photon energy of 944 eV as an example to illustrate
the effects of resonance absorption on the evolution dynamics of the fractions of
different charge states and of the population fractions of the SCH and DCH states,
which are shown in Figs. 2 and 3 for an x-ray pulse with peak intensity
$I_0$=2$\times$10$^{17}$ W/cm$^2$ and duration $\tau$=50 fs at a photon energy of 944 eV.
To obtain the result, we set a bandwidth $\Gamma$ to be 3 eV.

X-ray pulse with an energy of 944 eV can photoionize one $1s$ electron of
Ne-Ne$^{2+}$ to form SCH states of Ne$^{1+}$-Ne$^{3+}$, yet it cannot ionize
$1s$ electron of Ne$^{3+}$ as the photon energy is smaller than the
SIP of Ne$^{3+}$ (see table 1). However, there are inner-shell resonant absorption (IRA)
channels for Ne$^{3+}$. After the photoionization of $1s$ electron of Ne$^{2+}$,
SCH states belonging to the configuration $1s2s^22p^4$ of Ne$^{3+}$ is produced. Then
the DCH states $1s^02s^22p^5$ of Ne$^{3+}$ are effectively produced by the $1s\rightarrow 2p$
IRA process. The reason for the effectiveness of such processes lies in the large
absorption cross section near the resonances. The physical effects of IRA on the
evolution dynamics of populations can be seen in Fig. 2. The results have been averaged
over the spatial distribution of the light spot. The detailed method of averaging
over the space can be found in our work \cite{Gao4}. To save space, the dependence of
fraction evolution on the space is not explicitly given here. From the inspection
of Fig. 2, one can see that the IRA effects do not affect the fraction evolution
of Ne$^{1+}$ and Ne$^{2+}$, yet they have strong effects on higher ionization
stages, in particular for Ne$^{4+}$-Ne$^{5+}$. With the inclusion of the resonance
channels in the rate equation, the fraction of Ne$^{4+}$ is dramatically decreased,
while that of Ne$^{5+}$ is greatly enhanced. Such a decrease for Ne$^{4+}$ and
increase for Ne$^{5+}$ begin at -75 fs and the relative difference becomes larger
with the increase of time. The zero point of time is chosen at the moment of
peak intensity of x-ray laser. At the time of 0 and 50 fs, the fraction of Ne$^{4+}$
is decreased by 47\% and 55\% and that of Ne$^{5+}$ is increased by 135\% and
93\%, respectively. The fractions of Ne$^{6+}$-Ne$^{7+}$ are also evidently
enhanced with higher relative difference. Such a conclusion is easy to understand.
The DCH states of Ne$^{3+}$ decay dominantly to the SCH states of Ne$^{4+}$ by Auger
processes, which further decay to Ne$^{5+}$.

The effects of IRA processes can further be seen from the time evolution of the fractions
of DCH and SCH states produced in the interaction, which are shown in Fig. 3.
Without inclusion of IRA channels, the populations of DCH states of neon
charge states are very small ($\sim$10$^{-9}$) and thus they are not
shown in panel (a). Hence we can safely say that the DCH states at a photon energy of
944 eV are dominantly due to the resonance absorption channels from the SCH states
of Ne$^{3+}$. In panel (b), the populations of SCH states of
Ne$^{5+}$-Ne$^{7+}$ without IRA effects are not shown for the same reason.
For SCH states, there is little difference between the populations of Ne$^{1+}$-Ne$^{3+}$
with and without the consideration of IRA processes. However, there are large
discrepancies for the populations of higher charge states. For Ne$^{4+}$, the peak value
of the population (at -30fs) with IRA effects is about 50 times larger than the
corresponding value of without IRA effects. The SCH populations of Ne$^{5+}$-Ne$^{7+}$
with IRA effects (relative difference) are even much larger than the corresponding
values without IRA effects. These discrepancies
originate from the formation of DCH states of Ne$^{3+}$ and the followed Auger decay
to SCH states of Ne$^{4+}$. The large discrepancies of the
populations of DCH and SCH with and without IRA effects lead to the discrepancies
on the spectra accordingly.

From the above discussion, we know that the IRA processes have profound effects on
the time evolution of populations, especially for the DCH states. Without
the opening of the resonance absorption channels, the populations
of the DCH states are negligibly small. The populations will be enhanced by more than
six orders of magnitudes if we take the IRA processes into account.
To have a more complete picture of the CSD, we show it in Fig. 4 after spatially
and temporally averaging. The effects of IRA processes can readily be seen
from this figure. Overall, including IRA effects, the fraction of Ne$^{4+}$
is decreased by 49\%, while the fractions of Ne$^{5+}$-Ne$^{7+}$ are increased
by 114\%, 148\%, 180\%, respectively.

The prominent effects of IRA processes on the populations tell us that we must
include them in the rate equation to obtain accurate results. In the following,
these effects have been considered and therefore we do not state it again.
As an example, Fig. 5 shows the time evolution of emissivity of neon interacting
with an x-ray pulse (photon energy 944 eV) with a peak intensity of
$I_0$=2$\times$10$^{17}$ W/cm$^2$ and duration $\tau$=50 fs at time -50, 0, and
50 fs. In general, the emissivity decreases rapidly from t=-50 fs to t=0 fs
and then decreases smoothly from t=0 to t=50 fs. Two distinctive features
appear in the photon energy ranges of 830-930 eV and 930-1030 eV with the former
one produced by the SCH states and the latter one by DCH states.
The structures at lower energy range are dominantly contributed by Ne$^{1+}$-Ne$^{4+}$,
while at higher energy range only Ne$^{3+}$ contributes the largest emissivity.

After the irradiation of x-ray laser pulses, the emissivity of the highly
transient plasma is shown in Fig. 6 by averaging over the whole space and
time region. The dramatic effects due to IRA processes can readily be seen
from the comparison of the results with and without the IRA channels being
included. Without the opening of IRA channels, only the SCH states of
Ne$^{1+}$-Ne$^{3+}$ contribute to the emission spectra and the SCH states
of Ne$^{4+}$-Ne$^{5+}$ have negligibly small contribution. More clearly,
the emission features of DCH states above photon energy of 930 eV are
nearly completely invisible. The high efficiency of resonance absorption
of $1s\rightarrow2p$ for the production of DCH states can be seen from this
figure. We can also observe that the emission spectra of SCH and DCH states
have different characteristics. The quasi-continuous band of the DCH states
is broader than that of SCH ones for the same charge state.

In the above, we investigated the dynamical evolution of the CSD
and emissivity of transient plasmas produced in the interaction of neon
with x-ray laser pulses at a photon energy of 944 eV. The overall emission
features due to SCH and DCH states are discussed. In the following, we
focus our attention to the DCH spectroscopy. From Fig. 1, we know that
the lowest photon energy to produce DCH states is 937 eV, at which the
DCH states are created by the $1s\rightarrow2p$ resonance absorption,
just as we have just discussed in the above for h$\nu_0$=944 eV. We study the emissivity at two
different photon energy ranges of 930-1030 eV and above 1362 eV.
In the former region, the dominant mechanism of producing DCH states
is first photoionizing one $1s$ electron of neon ions
and then sequentially photoexciting another $1s$ electron
by $1s\rightarrow2p$ resonance absorption. In the latter region, the radiation
field sequentially photoionizes two $1s$ electrons one by one and
then the DCH states are produced.

We first investigate the DCH spectroscopy in the former photon energy region
where the resonance absorption exists.
Figure 7 shows the emissivity at six typical photon energies of 937, 944, 955,
968, 980, and 990 eV, which are the central $1s\rightarrow2p$ resonance energy
from the SCH states of Ne$^{2+}$-Ne$^{7+}$, respectively. The emission spectra at
944 eV have just been discussed detailedly in the above. The x-ray pulses are
assumed to have a peak intensity of 2$\times$10$^{17}$ W/cm$^2$
and duration of 50 fs in all cases. From the inspection of panels
(a)-(e) of Fig. 7, we can see evident DCH emissions from
Ne$^{2+}$-Ne$^{7+}$ ions. In general, the emissivity increases with the increase
of the photon energy except for that at 968 eV. The common feature for all six cases is that
prominent emission occurs at only around the resonance energies. This is
in agreement with the production mechanism discussed in the above.
From the IP of the DCH states for Ne-Ne$^{9+}$ ions
shown in table 1, we know that sequential photoionization of two $1s$ electrons
can only be accessible for Ne$^{2+}$ at photon energy higher than 986 eV.
This is because the photonionization
channels of $1s$ electrons from SCH states $1s2s^22p^6$ of Ne$^{1+}$
are opened at 990 eV and therefore the DCH
state $1s^02s^22p^6$ of Ne$^{2+}$ is effectively produced.
Such a fact is reflected in the emission spectra at 990 eV (Fig. 7(f)).
Besides the dominant emission features around the photon energy of 990 eV,
a clean emission structure located at 937 eV appears in Fig. 7(f).

For all other five cases with photon energy lower than 986 eV shown in Fig. 7(a)-(e),
the structure at 937 eV does not appear. However, new emission characteristics
are observed besides the above common feature. At the photon energy of
944 and 955 eV, the emission spectra are relatively pure and are dominantly due to
the $2p\rightarrow1s$ transitions from DCH states of Ne$^{3+}$
and Ne$^{4+}$. Yet for other cases, there are additional structures caused
by nearby ions. At 937 eV (resonance energy of Ne$^{2+}$), the emission
from Ne$^{3+}$ is even stronger than Ne$^{2+}$. At 968, 980, and 990 eV,
which corresponds to the central resonance energies of Ne$^{5+}$, Ne$^{6+}$,
and Ne$^{7+}$, respectively, emissions from nearby ions are evident in the emissivity.

How are the emissions from the nearby ions happened? In what follows,
we analyze the origin of these emission structures produced by nearby ions
by considering the results shown in Fig. 7(a) and 7(e) at 937 and 980 eV
as examples. First, we pay attention to the emissivity at 937 eV.
The emission line located at 937 eV originates from $2p\rightarrow 1s$
transitions from level $1s^02s^22p^6$ to $1s2s^22p^5$ of Ne$^{2+}$.
Around and lower than 937 eV, there are some structures except for that emitted by
Ne$^{2+}$. These emissions originate from the SCH states produced by the resonance
absorption $1s\rightarrow 3p$ from levels belonging to configuration
$1s^22s^22p^3$ of Ne$^{3+}$, as shown in table 2 (Nos. 1-9). Once
SCH states $1s2s^22p^33p$ are produced, they radiatively decay to the
levels of the ground configuration $1s^22s^22p^3$, resulting in the emissions
at photon energy range of lower than 937 eV. The strongest emission centered
at 944 eV originates from the DCH states of Ne$^{3+}$. It seems that DCH states
of Ne$^{3+}$ cannot be produced at 937 eV, which is the resonance energy of
Ne$^{2+}$. Intuitive feeling tells us that higher photon energy is needed to
create DCH states of Ne$^{3+}$. Yet there are channels for such a production
and the mechanism can be explained as follows. After the creation of DCH state
$1s^02s^22p^6$ for Ne$^{2+}$, it decays dominantly by Auger processes to SCH levels
belonging to the configurations of $1s2s^22p^4$, $1s2s2p^5$, and $1s2p^6$ of
Ne$^{3+}$. From levels of $1s2s2p^5$, there are $1s\rightarrow2p$ resonances
to produce DCH state $1s^02s2p^6$ of Ne$^{3+}$ (see Nos. 10-11 in table 2) with
resonance energy of $\sim$937.5 eV, which is very close to 937 eV. In such a way,
DCH state $1s^02s2p^6$ of Ne$^{3+}$ are produced. Then it decays to the lower levels
of $1s2s2p^5$ radiatively to emit at $\sim$944 eV (transitions Nos. 12-13 in table 2).
Due to the fine structure splitting of $1s2s2p^5$, the resonance energy is located
at $\sim$937 eV for those levels with higher energies, while it is located
at $\sim$944 eV for those levels with lower energies. As the oscillator strengths
of transitions (Nos. 12-13 in table 2) are very large, resulting in the strong emission
at $\sim$944 eV.

Secondly, we turn to the emissivity at photon energy of 980 eV ($1s\rightarrow2p$
resonance energy of Ne$^{6+}$), which is shown in Fig. 7(e). DCH spectroscopy of
Ne$^{5+}$ and Ne$^{6+}$ is evident in the figure, yet the production mechanism of DCH states
is different from that at 937 eV. From the inspection of table 1, we know that
photons with energy of 980 eV can ionize one $1s$ electron of Ne$^{3+}$,
yet they do not have enough energy to ionize one $1s$ electron of Ne$^{4+}$
and higher charge states. Hence it is impossible to produce DCH states
of Ne$^{6+}$ by first creation of $1s2s^m2p^n$ ($m+n$=3) of Ne$^{6+}$ from single
$1s$ photoionization and then
produce DCH states by $1s\rightarrow2p$ resonance absorption. Because of this,
there is no emission from Ne$^{6+}$ at 980 eV. Then how are
the DCH states of Ne$^{5+}$ and Ne$^{6+}$ produced? What kind of K-shell
hollow states are they? As we just discussed, photons with energy of 980 eV
can ionize one $1s$ electron up to Ne$^{3+}$ forming the SCH states
(dominated by $1s2s^22p^3$, $1s2s2p^4$, and $1s2p^5$) of Ne$^{4+}$,
which mainly decay to levels of $1s^22s^22p$, $1s^22s2p^2$ and $1s^22p^3$ of
Ne$^{5+}$. From these levels, there are $1s\rightarrow3p$ resonance
absorption channels of Ne$^{5+}$ to form SCH states of $1s2s^m2p^n3p$
($m+n$=3), which can further be photoexcited by $1s\rightarrow2p$ resonance
to form the DCH states $1s^02s^m2p^n3p$ ($m+n$=4) of Ne$^{5+}$.
To have a direct understanding of these resonance channels,
the absorption cross sections of the above two types of resonant transitions
are shown in Fig. 8(a). Many strong resonant absorptions are centered at
photon energy of 980 eV to form quasi-continuum bands. The channels can effectively
produce the DCH states of Ne$^{5+}$ and result in the emission around 980 eV.
Once DCH states $1s^02s^m2p^n3p$ ($m+n$=4) of Ne$^{5+}$ are produced,
they decay by Auger processes dominantly to levels of $1s2s^m2p^n3p$ ($m+n$=2) of
Ne$^{6+}$. From these levels, there are also many $1s\rightarrow2p$ resonance
absorption with the strongest peak being at 990 eV. Although the resonances
at 980 eV are weak compared with the stronger ones, the absorption cross section
(quasi-continuum band around 980 eV) is far larger than the direct photoionization
cross section of $2s$, $2p$, and $3p$ electrons and therefore create effective
channels to form DCH states of $1s^02s^m2p^n3p$ (m+n=3) of Ne$^{6+}$.
The above conclusion can be seen from Fig. 8(b), which shows the absorption
cross sections of $1s\rightarrow2p$ resonances from $1s2s^m2p^n3p$ of Ne$^{6+}$.
Once DCH states $1s^02s^m2p^n3p$ of Ne$^{6+}$ are produced, they decay
by radiative processes to the SCH states of $1s2s^m2p^n3p$ by
$2p\rightarrow1s$ centered at 990 eV. The emissions centered at $\sim$1004 eV
are mainly due to $3p\rightarrow1s$ transitions from SCH states of $1s2s^m2p^n3p$
of Ne$^{6+}$.

From the above discussion, we know that the production of DCH states of
Ne$^{5+}$ and Ne$^{6+}$ are rather complicated at photon energy 980 eV.
In particular, the DCH states of Ne$^{6+}$ originate from the weak resonances
and the DCH spectroscopy shows special features.
The emissivity is very sensitive to the laser intensity, as demonstrated
in Fig. 9, which shows the DCH spectroscopy at laser intensities of
$1\times$10$^{16}$, $3\times$10$^{16}$, $5\times$10$^{16}$, and
$1\times$10$^{17}$ W/cm$^2$. At the intensity of $1\times$10$^{16}$
W/cm$^2$, the dominant DCH emissions originate from Ne$^{5+}$
($2p\rightarrow1s$) centered at 980 eV, while the DCH spectroscopy from
Ne$^{6+}$ centered at 990 eV is nearly invisible. The emission structures
around 1004 eV originate from the SCH state transitions ($3p\rightarrow1s$)
to levels of $1s^22s^m2p^n$ (m+n=2) for Ne$^{6+}$.
With the increase of laser intensity, the DCH emissivity from Ne$^{6+}$
becomes larger and larger and the peak value exceeds that of Ne$^{5+}$
at the intensity of $\sim$6$\times$10$^{16}$ W/cm$^2$.

In the above, we investigated in detail the DCH spectroscopy at photon energy region
where there exist strong resonance absorptions of $1s\rightarrow2p$ from the SCH
states of different ionization stages of neon.
In what follows, we turn to the DCH spectroscopy at photon energy range of larger than
1362 eV, where no resonance absorption exists.
Figure 10 shows the DCH spectroscopy at photon energies of 1400, 1600, 1800, and
2000 eV. The laser intensity and duration are taken to be the same values as in the above
($I_0$=2$\times$10$^{17}$ W/cm$^2$ and $\tau$=50 fs). For x-rays in this photon energy range,
the creation mechanism of DCH states is simple. By sequentially absorbing two photons,
the DCH states of different ionization stages can be produced.
With the increase of photon energy,
the emissivity decreases from Fig. 10(a)-10(d). At lower photon energy,
the photoionization cross section is larger and therefore it is more efficient
to photoionize neon ions than at higher energy. The fact of larger cross section
at lower photon energy also explains the reason that the emissivity from the
DCH states of higher ionization stages are stronger than at higher energy.
In all cases, the emissions from DCH states of Ne$^{2+}$-Ne$^{6+}$ are weak,
while they increase with the increase of x-ray photon energy.

In this work, XFELs are assumed to have short temporal coherence
and the coherence effects are neglected. At present day, XFELs work at
a mode of self-amplified spontaneous emission and the longitudinal coherence
is not good \cite{Bostedt}. Recently, Ackermann {\it et al.} \cite{Ackermann}
reported the first seeded XFELs, which generated highly coherent source of radiation
with a very long temporal coherence. For such coherent XFELs, one has to consider the
quantum coherence effects in the dynamical excitation, ionization, and decay
processes of the produced transient plasmas. Li {\it et al.} \cite{Li} developed
a large scale quantum master equation approach to deal with such real dynamical
evolution driven by the coherent XFELs. The physical effects due to the longitudinal
coherence of XFELs are clearly demonstrated in this reference.

\section{Conclusion}

Neon interacting with ultraintense x-ray pulses is investigated by using
a time-dependent rate equation implemented in the detailed level accounting approximation.
The double core-hole (DCH) spectroscopy is investigated systematically for
x-ray pulses with different photon energy ranging from 937 eV to 2000 eV, which covers
two distinct regions with different features. The region with lower energy from
937 to 1030 eV, there exist many strong $1s\rightarrow 2p$ resonances from the SCH
states of Ne$^{2+}$-Ne$^{8+}$. Yet in another region with photon energy higher than 1362 eV
there are no resonance absorption for any ionization stage of neon.
In the resonance region, inner-shell resonant absorption (IRA) effects play important
roles on dynamical evolution of the level populations and emission properties of the highly
transient plasmas produced in the interaction, in particular for the DCH states.
The IRA effects are exemplified by studying the interaction of x-ray pulses
at a photon energy of 944 eV, which corresponds to the $1s\rightarrow 2p$ resonant
absorption from the SCH states ($1s2s^22p^4$, $1s2s2p^5$, and
$1s2p^6$) of Ne$^{3+}$. The population of DCH states is dramatically enhanced (more than
six orders of magnitudes) compared with that without the consideration of IRA.
The DCH spectroscopy at six typical photon energies of 937, 944, 955, 968, 980,
and 990 eV is investigated in detail. These photon energies correspond to the
$1s\rightarrow 2p$ resonance energies of Ne$^{2+}$-Ne$^{7+}$.
At 944 and 980 eV, the mechanisms of producing the DCH states are detailedly studied.
For x-ray pulses with photon energy larger than 1362 eV (where no resonances
absorption exist), the DCH spectroscopy shows completely different features
compared with that at 937-1030 eV. DCH states from all possible ionization stages of
Ne$^{2+}$-Ne$^{8+}$ can be produced by multiple photon ionization.
Yet the dominant DCH states are due to higher ionization stages of
Ne$^{7+}$-Ne$^{8+}$. With the increase of photon energy, the contributions from the lower
charge states Ne$^{2+}$-Ne$^{6+}$ become larger.

\ack
This work was supported by the National Natural Science Foundation of China
under Grant No. 11204376, No. 11274382 and No. 11274383.

\newpage

\begin{table}
\caption{\label{tab.tar} Single ionization potential (SIP) and double ionization potential (DIP) of
SCH and DCH states for Ne-Ne$^{9+}$ ions. The ground and single core-hole configurations
are given to aid the understanding. Theoretical results are given except for those
experimental values wherever available indicated after the ionization potentials.}
\begin{indented}
\item[]\begin{tabular}{@{}lllll}
\br
Ion & Ground config. & SIP (eV)  & SCH config. & DIP (eV) \\\hline
Ne        & $1s^22s^22p^6$ &  870.3\cite{Thomas2}  &              &        \\
Ne$^{1+}$ & $1s^22s^22p^5$ &  892.6                & $1s2s^22p^6$ & 986.0  \\
Ne$^{2+}$ & $1s^22s^22p^4$ &  923.1                & $1s2s^22p^5$ & 1021.0 \\
Ne$^{3+}$ & $1s^22s^22p^3$ &  957.9                & $1s2s^22p^4$ & 1061.2 \\
Ne$^{4+}$ & $1s^22s^22p^2$ & 1001.8\cite{Bruch}    & $1s2s^22p^3$ & 1106.1 \\
Ne$^{5+}$ & $1s^22s^22p$   & 1048.5\cite{Bruch}    & $1s2s^22p^2$ & 1152.3 \\
Ne$^{6+}$ & $1s^22s^2$     & 1099.1\cite{Bruch}    & $1s2s^22p$   & 1202.3 \\
Ne$^{7+}$ & $1s^22s$       & 1143.2                & $1s2s^2$     & 1253.6 \\
Ne$^{8+}$ & $1s^2$         & 1195.3\cite{NIST}     & $1s2s$       & 1311.6 \\
Ne$^{9+}$ & $1s$           & 1362.2                &              &        \\
\br
\end{tabular}
\end{indented}
\end{table}

\begin{table}
\caption{\label{tab.tar} Resonance energy (RE) (eV) and weighted oscillator strength (gf)
of $1s\rightarrow 3p$ ($1s^22s^22p^3\rightarrow1s2s^22p^33p$, Nos. 1-9) and $1s\rightarrow 2p$
($1s2s2p^5\rightarrow 1s^02s2p^6$, Nos. 10-13) transitions for the production of SCH and DCH
states of Ne$^{3+}$ near 937 and 944 eV. Only those with gf larger than 0.02
near 937 eV and those larger than 0.1 near 944 eV are given for simplicity.}
\begin{indented}
\item[]\begin{tabular}{@{}lllcc}
\br
No. & Lower level & Upper level & RE   & gf \\\hline
1 & $[1s^22s^22p_{1/2}(2p_{3/2}^2)_2]_{3/2}$ & $[((1s2s^22p_{1/2})_1(2p_{3/2}^2)_2)_13p_{1/2}]_{1/2}$ & 934.81 & 0.020 \\
2 & $[1s^22s^22p_{1/2}(2p_{3/2}^2)_2]_{3/2}$ & $[((1s2s^22p_{1/2})_1(2p_{3/2}^2)_2)_13p_{3/2}]_{5/2}$ & 934.82 & 0.047 \\
3 & $[1s^22s^22p_{1/2}(2p_{3/2}^2)_2]_{3/2}$ & $[((1s2s^22p_{1/2})_1(2p_{3/2}^2)_2)_13p_{1/2}]_{3/2}$ & 934.85 & 0.039 \\
4 & $[1s^22s^22p_{1/2}(2p_{3/2}^2)_2]_{5/2}$ & $[((1s2s^22p_{1/2})_1(2p_{3/2}^2)_2)_33p_{3/2}]_{5/2}$ & 930.88 & 0.107 \\
5 & $[1s^22s^22p_{1/2}(2p_{3/2}^2)_2]_{5/2}$ & $[((1s2s^22p_{1/2})_1(2p_{3/2}^2)_2)_23p_{3/2}]_{7/2}$ & 933.13 & 0.055 \\
6 & $[1s^22s^22p_{1/2}(2p_{3/2}^2)_2]_{3/2}$ & $[(1s2s^2(2p_{3/2}^3)_{3/2})_13p_{3/2}]_{3/2}$         & 930.95 & 0.067 \\
7 & $[1s^22s^22p_{1/2}(2p_{3/2}^2)_2]_{3/2}$ & $[((1s2s^22p_{1/2})_1(2p_{3/2}^2)_2)_23p_{1/2}]_{5/2}$ & 933.11 & 0.039 \\
8 & $[1s^22s^2(2p_{3/2}^3)_{3/2}]_{3/2}$     & $[((1s2s^22p_{1/2})_1(2p_{3/2}^2)_0)_13p_{3/2}]_{1/2}$ & 930.92 & 0.027 \\
9 & $[1s^22s^2(2p_{3/2}^3)_{3/2}]_{3/2}$     & $[(1s2s^2(2p_{3/2}^3)_{3/2})_13p_{3/2}]_{5/2}$         & 932.75 & 0.031 \\
10& $[(1s2s)_0(2p_{3/2}^3)_{3/2}]_{3/2}$     & $[1s^02s2p^6]_{1/2}$                                   & 937.58 & 0.088 \\
11& $[(1s2s)_1(2p_{3/2}^3)_{3/2}]_{1/2}$     & $[1s^02s2p^6]_{1/2}$                                   & 937.52 & 0.041 \\
12& $[(1s2s)_12p_{1/2}]_{3/2}$               & $[1s^02s2p^6]_{1/2}$                                   & 944.98 & 0.506 \\
13& $[(1s2s)_02p_{1/2}]_{1/2}$               & $[1s^02s2p^6]_{1/2}$                                   & 944.92 & 0.255 \\
\br
\end{tabular}
\end{indented}
\end{table}

\clearpage
\begin{figure}[htb]
\includegraphics*[width=5in]{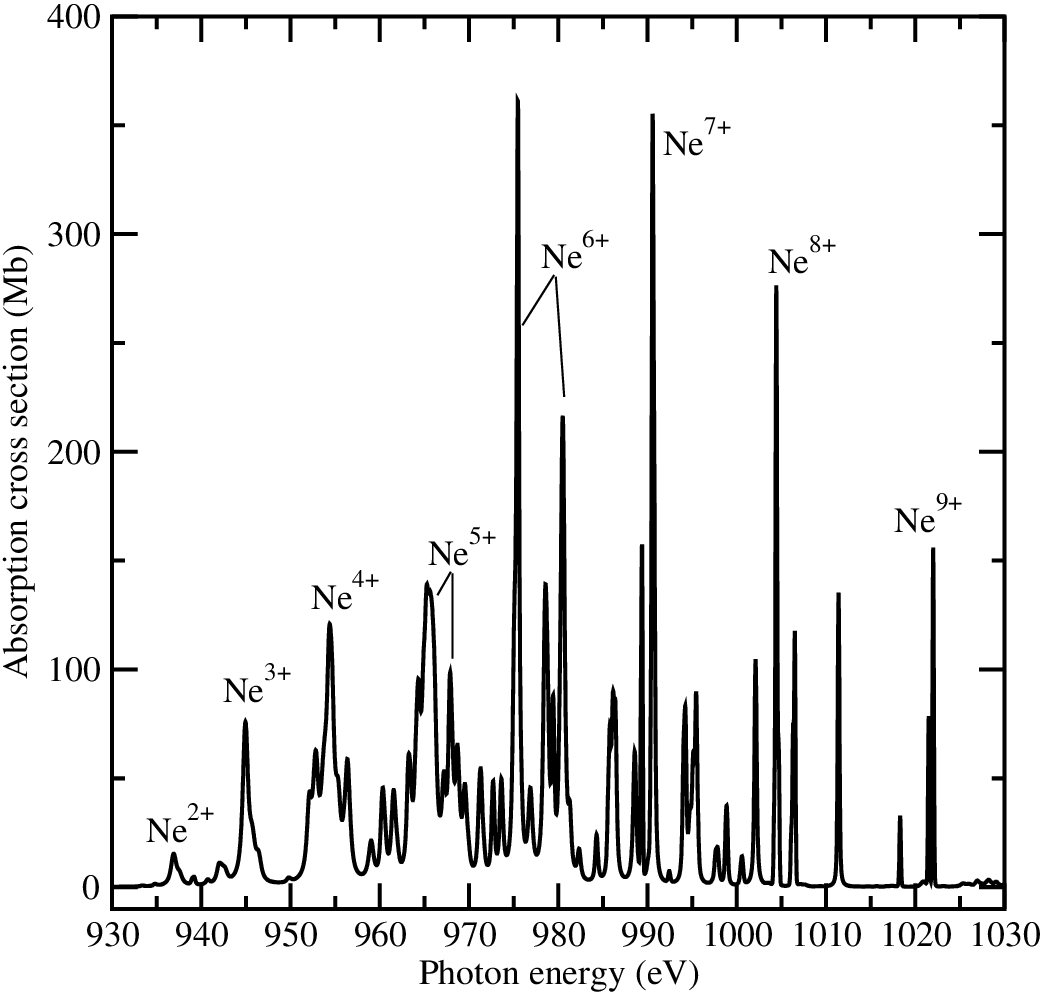}
\caption{Absorption cross section (in Mb=1.0$\times$10$^{-18}$ cm$^2$)
of $1s\rightarrow 2p$ resonance transitions from the SCH states $1s2s^m2p^n$
($m+n$=8-1) of neon charge states from Ne$^{2+}$ to Ne$^{8+}$.
Note that Ne$^{9+}$ has only one electron with the ground state of $1s$.}
\end{figure}

\clearpage
\begin{figure}[htb]
\includegraphics*[width=5in]{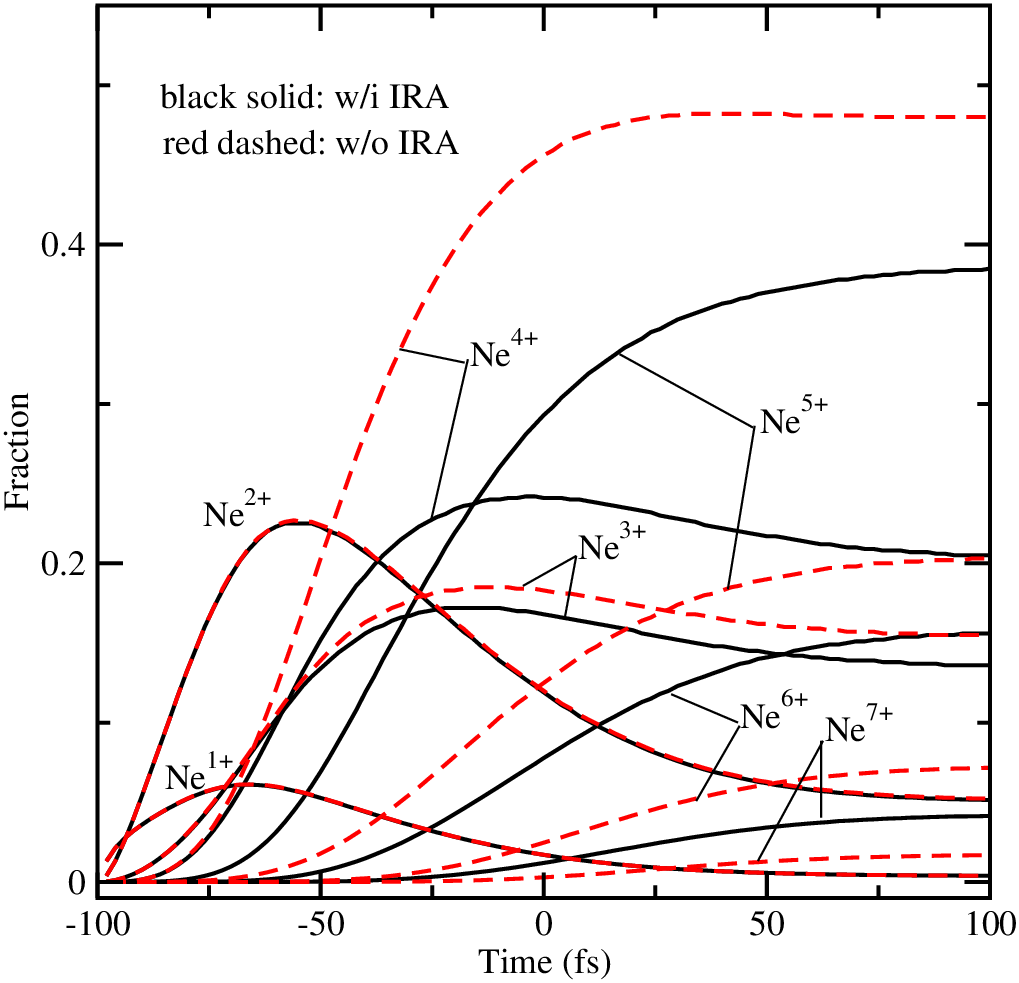}
\caption{(Color online) Fraction evolution of neon charge states
interacting with an x-ray pulse of peak intensity $I_0$=2$\times$10$^{17}$ W/cm$^2$,
duration $\tau$=50 fs with a central photon energy of 944 eV.
The black solid and red dashed lines represent the results with and without
IRA effects, respectively.}
\end{figure}

\clearpage
\begin{figure}[htb]
\includegraphics*[width=5in]{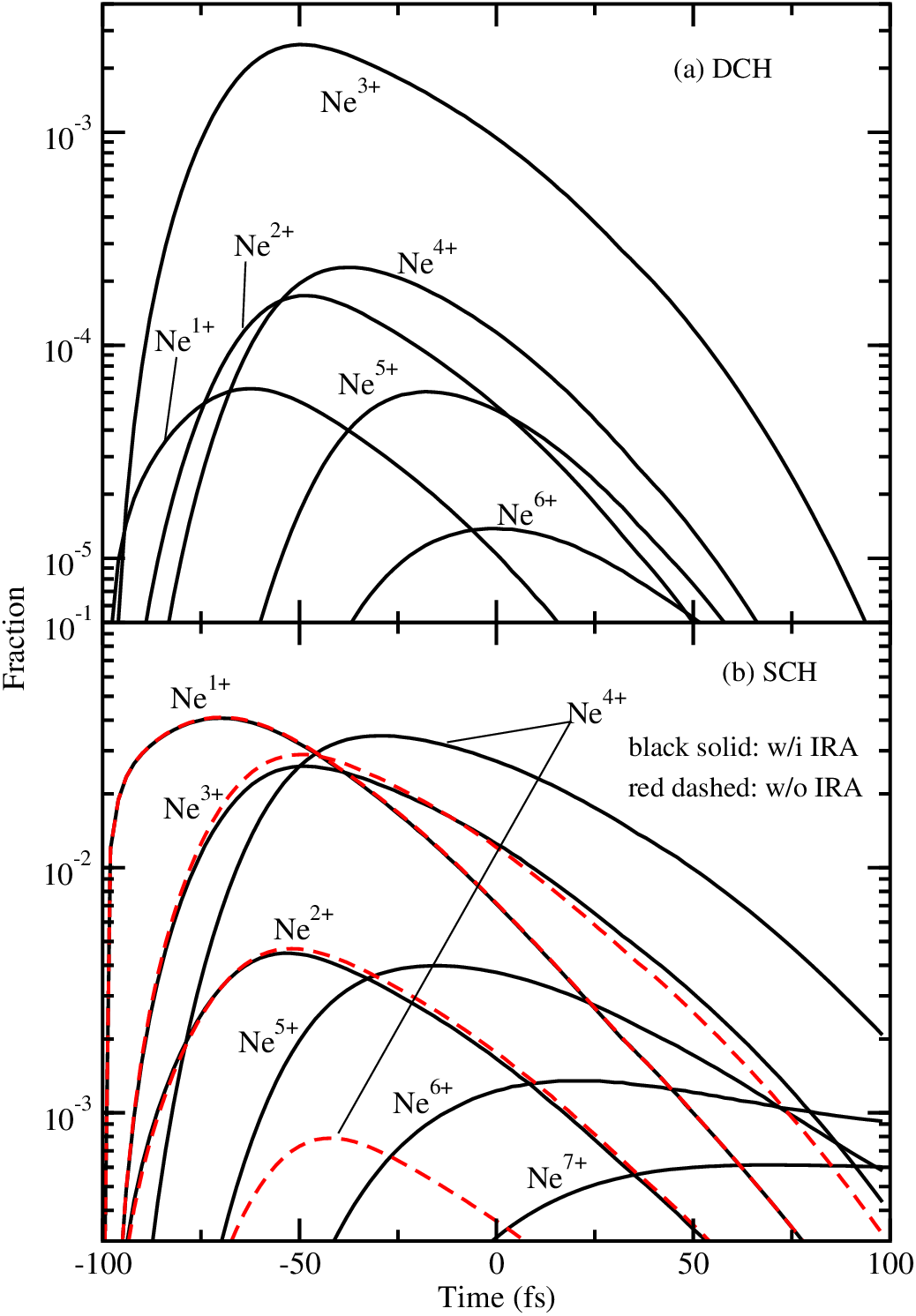}
\caption{(Color online) Fraction evolution of (a) DCH and
(b) SCH states of different neon charge states. The parameters of x-ray pulses
are the same as in Fig.2. The black solid and red dashed
lines represent the results with and without IRA effects, respectively.
In panel (a), the populations of DCH states without IRA effects are too small
to show. In panel (b), the populations of SCH states of Ne$^{5+}$-Ne$^{7+}$
without IRA effects are not shown for the same reason.}
\end{figure}

\clearpage
\begin{figure}[htb]
\includegraphics*[width=5in]{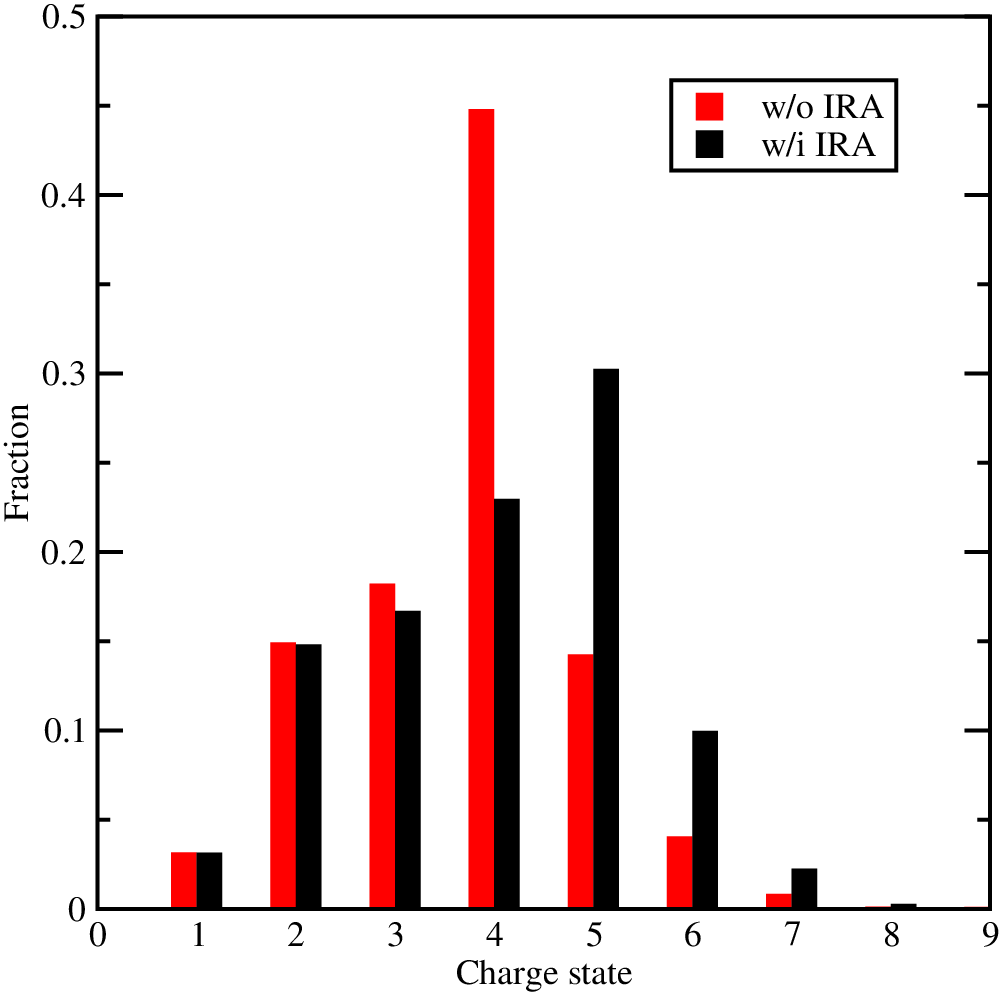}
\caption{(Color online) Charge state distribution of neon after spatial
and temporally averaging. The black and red bars represent the results with and without
IRA effects, respectively. The parameters of the laser pulse are the same
as in Fig. 2.}
\end{figure}

\clearpage
\begin{figure}[htb]
\includegraphics*[width=5in]{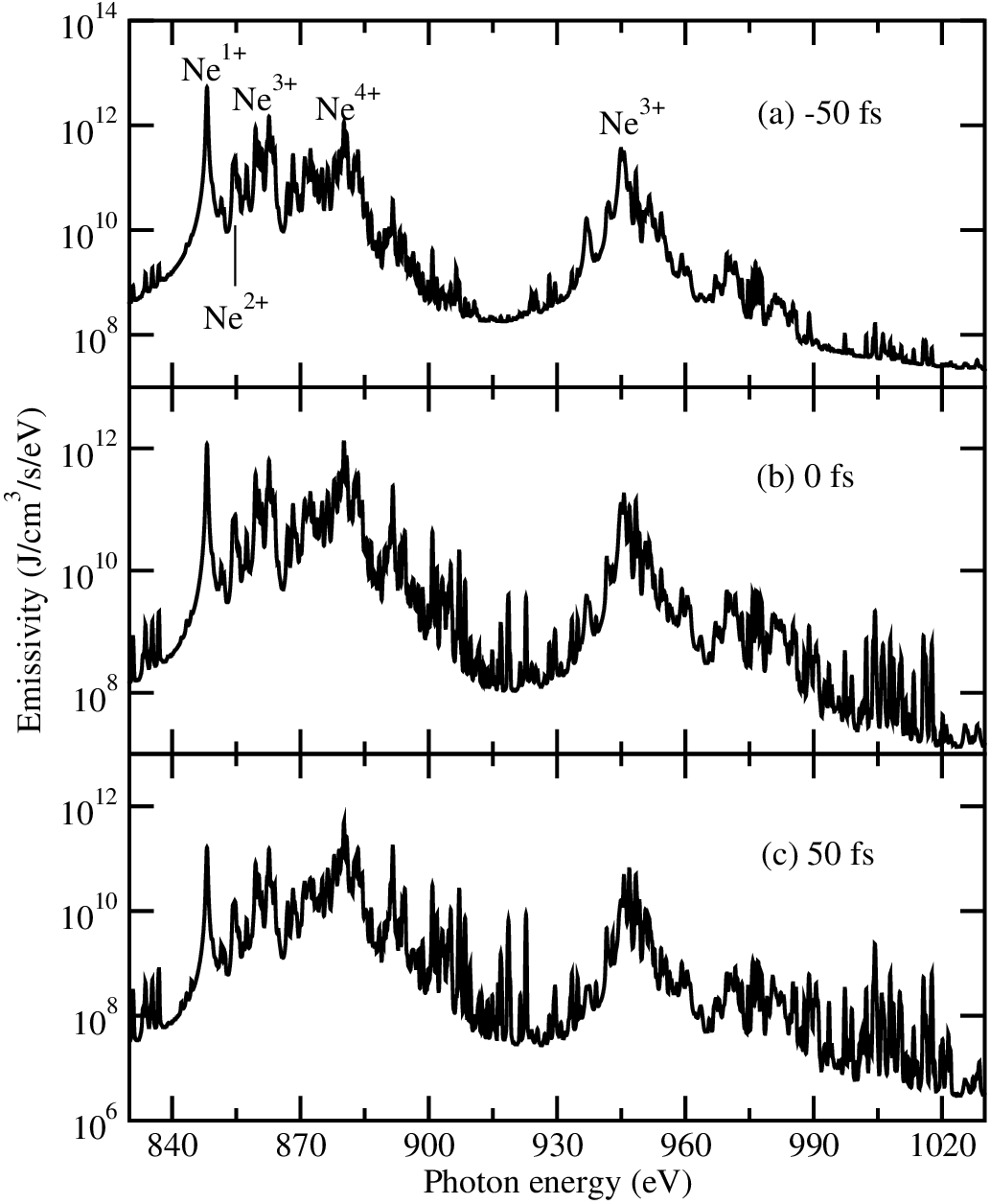}
\caption{Emission spectra of SCH and DCH states of neon at time
of (a) -50, (b) 0 and (c) 50 fs in the photon energy range of 830-1030 eV.
The parameters of the laser pulse are the same as in Fig. 2.
The structures at lower photon energy range ($<$930 eV) are due to the SCH
states and those at higher photon energy range ($>$930 eV) are dominantly
due to the DCH states.}
\end{figure}

\clearpage
\begin{figure}[htb]
\includegraphics*[width=5in]{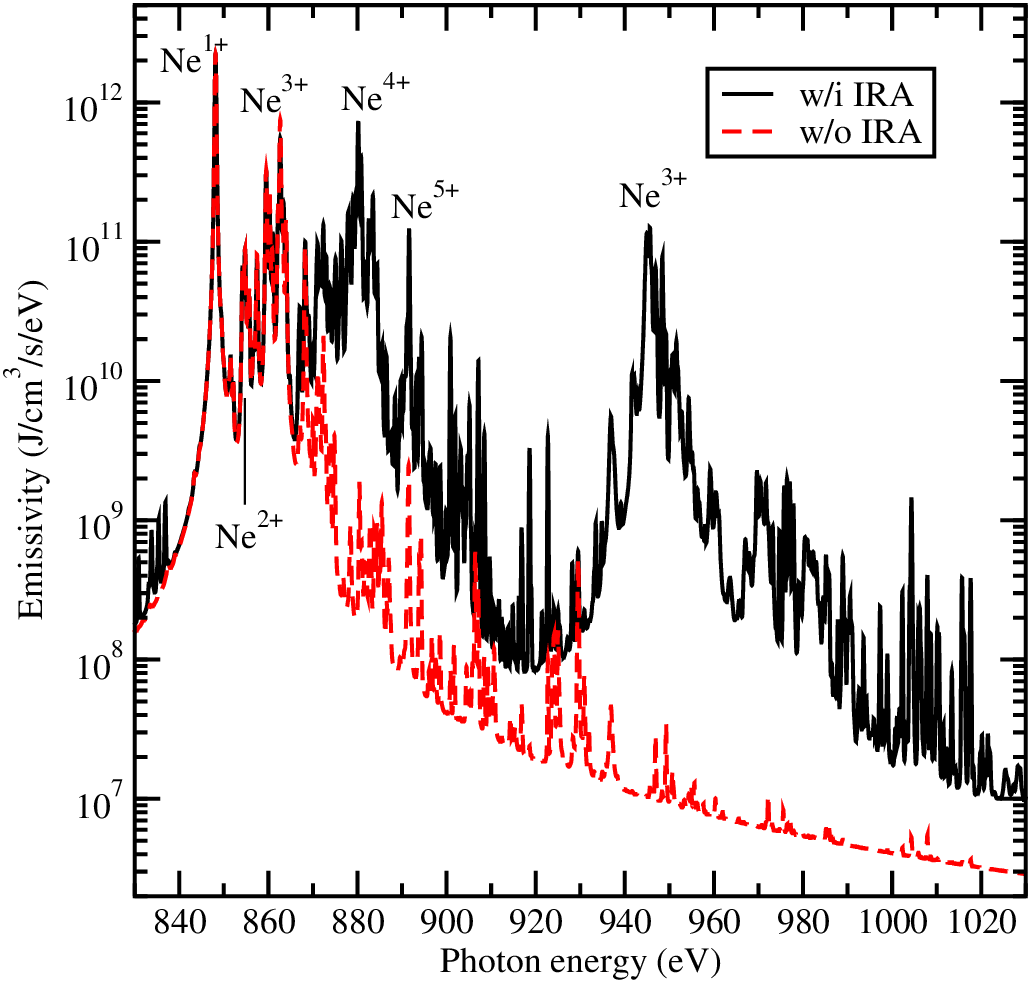}
\caption{(Color online) Comparison between the emission spectra of neon
with (black solid line) and without (red dashed line) IRA effects being considered
after spatial and temporally averaging.}
\end{figure}

\clearpage
\begin{figure}[htb]
\includegraphics*[width=5in]{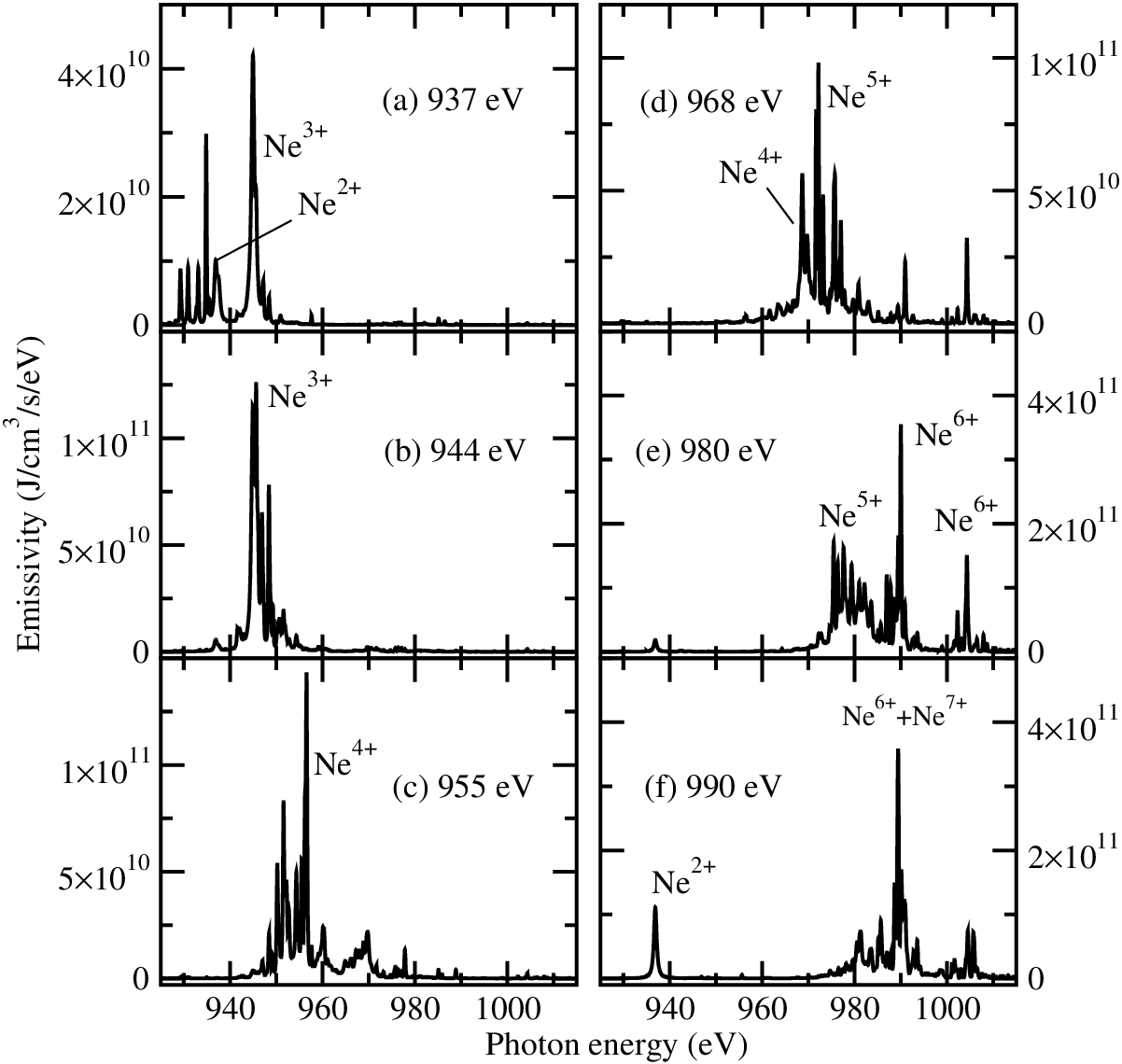}
\caption{DCH emission spectra of neon interacting with x-ray pulses
of peak intensity of 2$\times$10$^{17}$ W/cm$^2$, duration of 50 fs with photon
energies of (a) 937 eV, (b) 944 eV, (c) 955 eV, (d) 968 eV, (e) 980 eV
and (f) 990 eV, respectively.}
\end{figure}

\clearpage
\begin{figure}[htb]
\includegraphics*[width=5in]{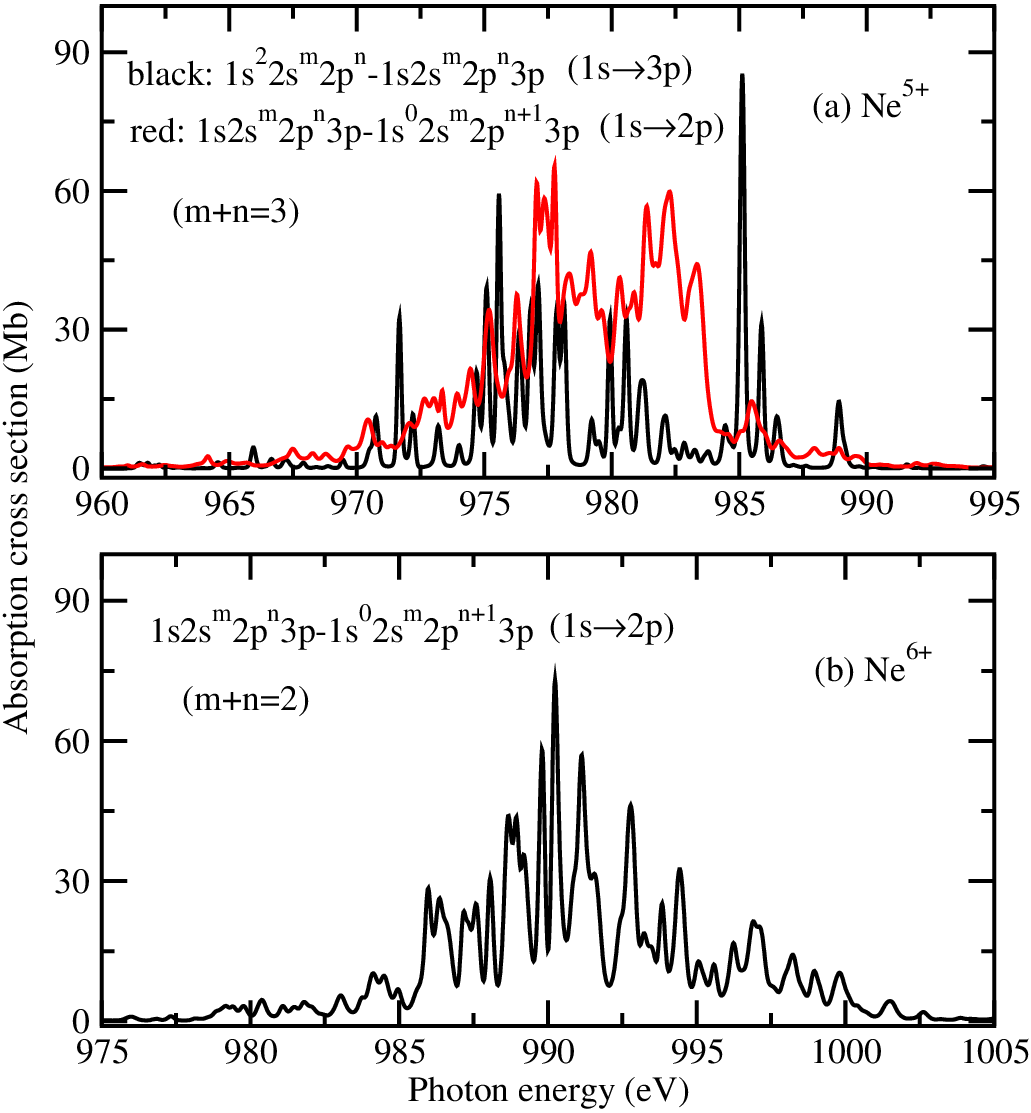}
\caption{(Color online) Absorption cross section of (a) Ne$^{5+}$
resonances of $1s^22s^m2p^n\rightarrow 1s2s^m2p^n3p$ and
$1s2s^m2p^n3p\rightarrow 1s^02s^m2p^{n+1}3p$ ($m+n$=3), and (b) Ne$^{6+}$
from $1s2s^m2p^n3p\rightarrow 1s^02s^m2p^{n+1}3p$ ($m+n$=2).}
\end{figure}

\clearpage
\begin{figure}[htb]
\includegraphics*[width=5in]{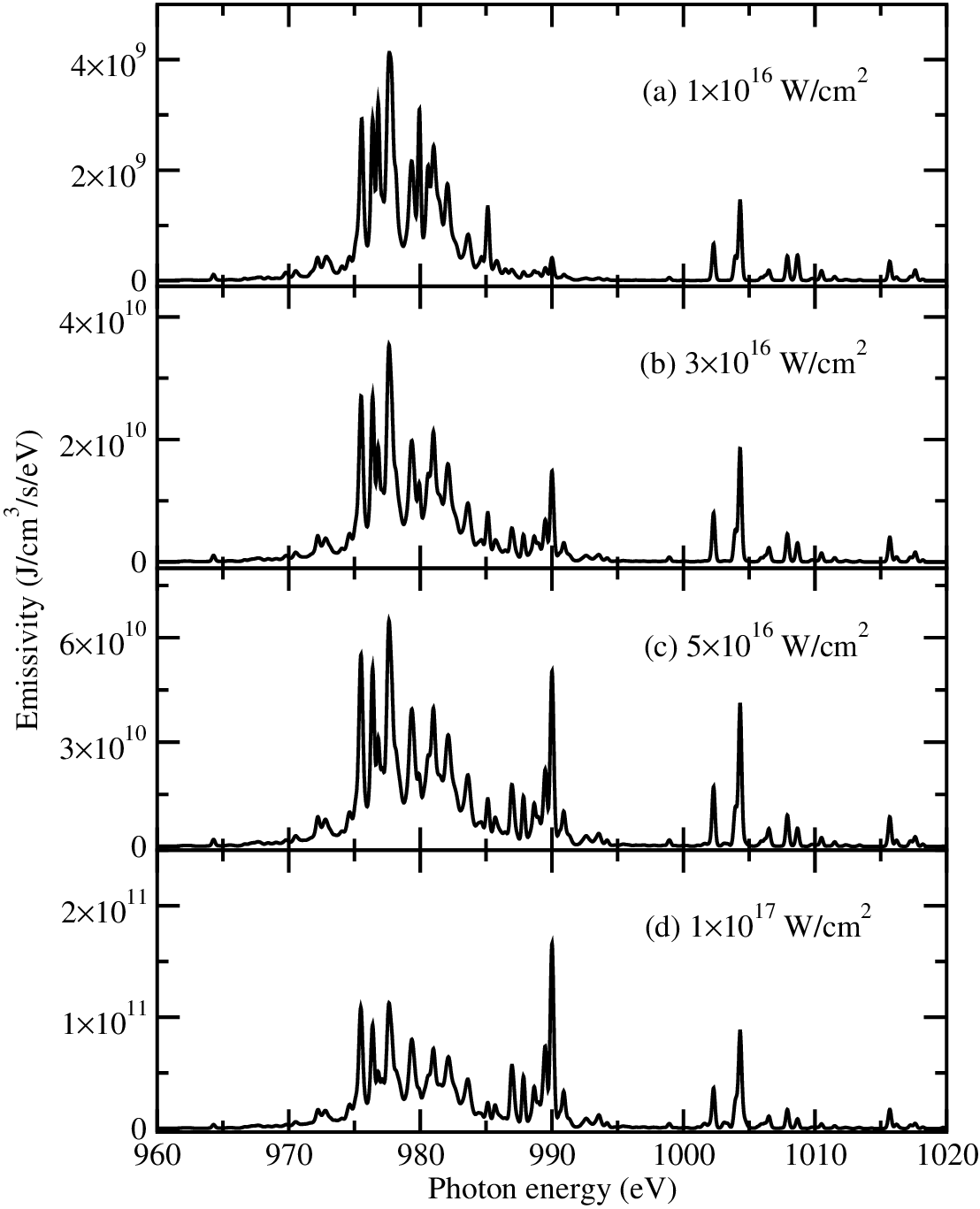}
\caption{DCH emission spectra of neon interacting with x-ray pulses
with photon energy of 980 eV, duration of 50 fs, and intensity of (a) 1$\times$10$^{16}$,
(b) 3$\times$10$^{16}$, (c) 5$\times$10$^{16}$ and (d) 1$\times$10$^{17}$ W/cm$^2$,
respectively.}
\end{figure}

\clearpage
\begin{figure}[htb]
\includegraphics*[width=5in]{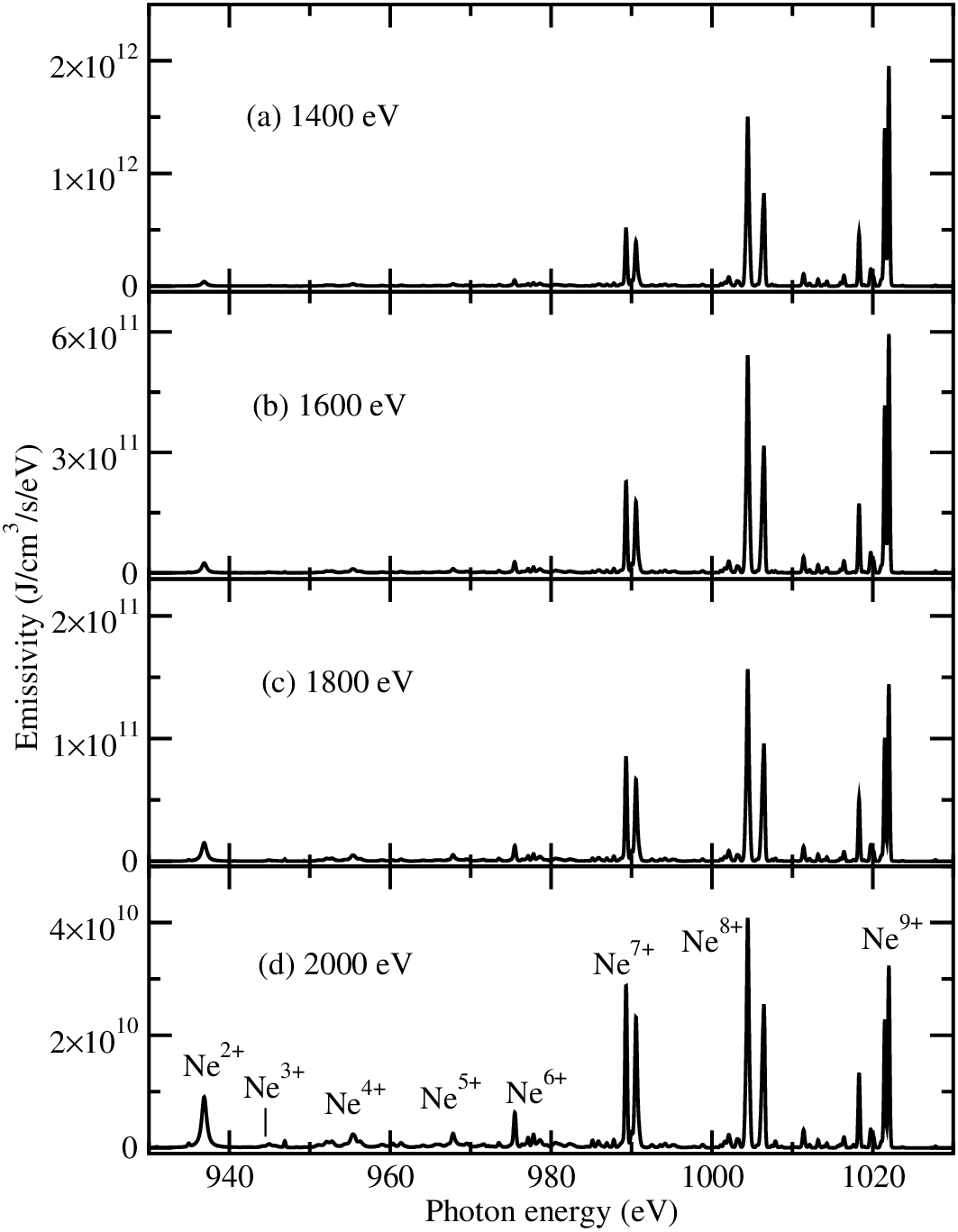}
\caption{DCH emission spectra of neon interacting with x-ray pulses
with photon energy of (a) 1400, (b) 1600, (c) 1800, and (d) 2000 eV.
The parameters of the laser pulse are the same as in Fig. 2.}
\end{figure}

\end{document}